\documentclass[preprint,preprintnumbers,amsmath,amssymb]{revtex4}
\usepackage{graphicx,epsfig}
\usepackage{dcolumn}
\usepackage{bm}
\begin{document}
\title{Orientational relaxation in a discotic liquid crystal}
\author{Dwaipayan Chakrabarti, Biman Jana, and Biman Bagchi\footnote[2] 
{For correspondence: bbagchi@sscu.iisc.ernet.in}}
\affiliation{Solid State and Structural Chemistry Unit, Indian Institute of Science, Bangalore 560012, India}

\date \today

\begin{abstract}

We investigate orientational relaxation of a model discotic liquid crystal, 
consists of disc-like molecules, by molecular dynamics simulations along two isobars starting 
from the high temperature isotropic phase. The two isobars have been so chosen that (A) the phase 
sequence isotropic (I)-nematic (N)-columnar (C) appears upon cooling along one of them
and (B) the sequence isotropic (I)-columnar (C) along the other. While the orientational 
relaxation in the isotropic phase near the I-N phase transition in system (A) shows a power law decay at 
short to intermediate times, such power law relaxation is {\it not} observed in the 
isotropic phase near the I-C phase boundary in system (B). 
In order to understand this difference (the existence or the
absence of the power law decay), we calculated the
the growth of the orientational pair distribution functions (OPDF) near the I-N phase boundary
and also near the I-C phase boundary. 
We find that OPDF shows a marked growth in long range correlation 
as the I-N phase boundary is approached in the I-N-C system (A),
but such a growth is absent in the I-C system, which appears to be consistent with the 
result that I-N phase transition in the former is weakly first order while the 
the I-C phase transition in the later is not weak.  As 
the system settles into the nematic phase, the decay of the single-particle second-rank 
orientational OTCF follows a pattern that is similar to what is observed with calamitic 
liquid crystals and supercooled molecular liquids.   
 
\end{abstract}

\maketitle

\section{Introduction}

The anisotropy in molecular shape plays a crucial role in the rich phase behavior 
that thermotropic liquid crystals exhibit \cite{deGennes-book,Chandrasekhar-book}. 
Calamitic liquid crystals, that comprise rodlike molecules, are long known and 
their phase behavior and dynamics have been extensively investigated over 
decades \cite{deGennes-book,Chandrasekhar-book}. The discovery of discotic liquid 
crystals, that consist of disclike molecules, is, however, more recent and dates 
back only to the late 1970s \cite{Chandrasekhar-Pramana-1970}. Due to its unique 
structural, elctrical and optical properties, discotic liquid crystals have drawn 
considerable amount of interests in recent past \cite{Ohta-Proc-1999,
Adam-Nature-1994,Yoshino-Proc-1999}. Upon cooling from 
the high temperature isotropic (I) phase, discotic liquid crystals typically 
exhibit a nematic (N) phase and/or a columnar (C) phase 
\cite{Chandrasekhar-Ranganath-RPP-1990}. The discotic nematic phase is 
analogous to the nematic phase formed by rodlike molecules in that there is a 
long-range orientational order without the involvement of any long-range 
translational order. In the columnar phase that is typical of discotic liquid 
crystals, the molecules are stacked on top of each other giving rise to a columnar
structure. These columns form a long-range two dimensional order in the 
orthogonal plane with either a hexagonal or a rectangular symmetry. While the 
sequence of phases I-N-C has been observed experimentally with a number of 
discotic liquid crystals upon cooling, there have been only a few cases where I-C 
transition is observed \cite{Hindmarsh-JMC-1993}. Although these are highly interesting 
systems to study and computer simulation studies of model liquid crystals have undergone 
an upsurge in recent times \cite{Pasini-Zannoni-book,Wilson-IRPC-2005}, we are aware of only 
very few studies of orientational relaxation on discotic liquid crystals \cite{Zannoni-JMC-2001} 
compared to that on rod-like molecules. 

Discotic molecules typically contain an aromatic core with flexible chains added in the 
equatorial plane \cite{Zannoni-JMC-2001}. While atomistic models could in 
principle be undertaken, molecular models, where mesogens are approximated as 
particles with well-defined anisotropic shape, find their utility in obtaining a 
rather generalized view. A simple approach along this line involves consideration 
of purely repulsive models involving hard bodies \cite{Allen-ACP-1993}. This 
rather extreme choice is inspired by the idea that the equilibrium structure of a 
dense liquid is essentially determined by the repulsive forces which fix the 
molecular shape \cite{Andersen-ACP-1976}. Along this line, thin hard platelets 
\cite{Eppenga-Frenkel-MP-1984}, hard oblate ellipsoids of revolution 
\cite{Frenkel-Mulder-MP-1985,Allen-PRL-1990}, and cut hard spheres 
\cite{Veerman-Frenkel-PRA-1992} have been investigated. Such an approach is
appealing for its simplicity \cite{Allen-ACP-1993}. However, temperature plays no 
direct role in purely repulsive models on the contrary to what is desired for
thermotropic liquid crystals \cite{Allen-ACP-1993}. In this respect, the Gay-Berne 
pair potential \cite{Gay-Berne-JCP-1981}, which is essentially a generalization
of the Lennard-Jones potential to incorporate anisotropic interactions, or one of 
its variants \cite{Gay-Berne-JCP-1981}, where mesogens are approximated with soft 
ellipsoids of revolution, appears to serve as a more realistic model. In fact, discotic 
liquid crystals, modeled by the Gay-Berne family of potentials, have been found to 
capture the key features of the experimentally observed phase behavior 
\cite{Emerson-MP-1994,Bates-Luckhurst-JCP-1996,Capiron-PRE-2003}. In a density 
functional theoretical approach with a form of the Gay-Berne potential modified for 
discotic liquid crystals, the  isotropic-nematic-columnar phase behavior has recently 
been studied for various aspect ratios \cite{Coussaert-Baus-JCP-2002}. 

Dynamics of discotic liquid crystals have drawn attention as well 
\cite{Allen-PRL-1990,Maliniak-JCP-1992,Zamir-JACS-1994,Groothues-LC-1995,
Dong-MP-1999,Dong-Morcombe-LC-2000,Dvinskikh-PRE-2002,Mulder-JACS-2003,
Zhang-Dong-PRE-2006}. The focus has often been on the dynamics of the columnar
phase \cite{Zamir-JACS-1994,Dong-MP-1999,Dvinskikh-PRE-2002,Mulder-JACS-2003,
Zhang-Dong-PRE-2006}. In this work, we have undertaken molecular dynamics 
simulations of a system of oblate ellipsoids of revolution interacting with a 
modified Gay-Berne pair potential to study temperature dependent orientational 
relaxation along two isobars. We have chosen two isobars such that the phase sequence 
I-N-C appears upon cooling along the one at a higher pressure and the sequence I-C along 
the other. We have investigated temperature dependent orientational relaxation across 
the isotropic-nematic transition and in the isotropic phase near the I-C phase 
boundary with a focus on the short-to-intermediate time decay behavior. This work 
follows up our recent work \cite{Chakrabarti-PRL-2005}, which has reported the 
emergence of power law decay regime(s) in orientational relaxation across the 
isotropic-nematic transition. In the spirit of the universal power law in 
orientational relaxation in thermotropic liquid crystals suggested therein 
\cite{Chakrabarti-PRL-2005}, we compare the orientational dynamics we observed here 
with those of calamitic liquid crystals obtained from recent optical Kerr effect 
measurements \cite{Gottkea-JCP-2002,Gottkeb-JCP-2002,Cang-CPL-2002,Li-JPCB-2005} and 
molecular dynamics simulations studies \cite{Jose-Bagchi-JCP-2004,Chakrabarti-PRL-2005,
Bertolini-JPCB-2005}. We further discuss the analogous dynamics observed in supercooled 
molecular liquids.   

The rest of the paper is organized as follows. Section II describes the model we 
have studied here along with some simulation details. In section III, we present 
the results with discussion. Section IV presents a theoretical analysis of the origin of 
the power law decay. Section V discusses about orientational pair distribution function 
before we conclude in section VI with a summary of the results and a few comments.

\section{Model and details of the simulation}

The Gay-Berne (GB) pair potential, where each ellipsoid of revolution has a 
single-site representation, is an elegant generalization of the extensively used 
isotropic Lennard-Jones potential to incorporate anisotropy in both the attractive
and the repulsive parts of the interaction \cite{Gay-Berne-JCP-1981}. In the GB 
pair potential, $i$th ellipsoid of revolution is represented by the position 
${\bf r}_{i}$ of its center of mass and a unit vector ${\bf e}_{i}$ along the 
short axis in the case of an oblate. In this work, we have employed the form of 
the GB potential that has been modified by Bates and Luckhurst for discotic liquid 
crystals \cite{Bates-Luckhurst-JCP-1996}. In this modified form, the interaction 
between two oblate ellipsoids of revolution $i$ and $j$ is given by
\begin{equation}
U_{ij}^{GB}({\bf r}_{ij},{\bf e}_{i},{\bf e}_{j})  =
4\epsilon({\bf \hat r}_{ij},{\bf e}_{i},{\bf e}_{j})(\rho_{ij}^{-12} -
\rho_{ij}^{-6})
\end{equation}
where
\begin{equation}
\rho_{ij} = \frac{r_{ij} - \sigma({\bf \hat r}_{ij},{\bf e}_{i},{\bf e}_{j})
+ \sigma_{ff}}{\sigma_{ff}}.
\end{equation}
Here $\sigma_{ff}$ defines the thickness or equivalently, the separation between 
the two in a face-to-face configuration, $r_{ij}$ is the distance between the 
centers of mass of the ellipsoids of revolution $i$ and $j$, and 
${\bf \hat r}_{ij} = {\bf r}_{ij} / r_{ij} $ is a unit vector along the
intermolecular separation vector ${\bf r}_{ij}$. The molecular shape
parameter $\sigma$ and the energy parameter $\epsilon$ both depend on the
unit vectors ${\bf e}_{i}$ and ${\bf e}_{j}$ as well as on
${\bf \hat r}_{ij}$ as given by the following set of equations:
\begin{equation}
\sigma({\bf \hat r}_{ij},{\bf e}_{i},{\bf e}_{j}) = \sigma_{0}\left[1 -
\frac{\chi}{2} \left\{\frac{({\bf e}_{i}\cdot{\bf \hat r}_{ij} +
{\bf e}_{j}\cdot{\bf \hat r}_{ij} )^{2}}
{1 + \chi({\bf e}_{i}\cdot{\bf e}_{j})} +
\frac{({\bf e}_{i}\cdot{\bf \hat r}_{ij} -
{\bf e}_{j}\cdot{\bf \hat r}_{ij})^{2}}{1 - \chi({\bf e}_{i} \cdot
{\bf e}_{j})}\right\}\right]^{-1/2}
\end{equation}
with $\chi = (\kappa^{2} - 1) / (\kappa^{2} + 1)$ and
\begin{equation}
\epsilon({\bf \hat r}_{ij},{\bf e}_{i},{\bf e}_{j}) = \epsilon_{0}
[\epsilon_{1}({\bf e}_{i},{\bf e}_{j})]^{\nu}
[\epsilon_{2}({\bf \hat r}_{ij},{\bf e}_{i},{\bf e}_{j})]^{\mu}
\end{equation}
where the exponents $\mu$ and $\nu$ are adjustable, and
\begin{equation}
\epsilon_{1}({\bf e}_{i},{\bf e}_{j}) =
[1 - \chi^{2}({\bf e}_{i}\cdot{\bf e}_{j})^{2}]^{-1/2}
\end{equation}
and
\begin{equation}
\epsilon_{2}({\bf \hat r}_{ij},{\bf e}_{i},{\bf e}_{j}) = 1 -
\frac{\chi ^{\prime}}{2} \left[\frac{({\bf e}_{i}\cdot{\bf \hat r}_{ij} +
{\bf e}_{j}\cdot{\bf \hat r}_{ij} )^{2}}
{1 + \chi^{\prime}({\bf e}_{i}\cdot{\bf e}_{j})} +
\frac{({\bf e}_{i}\cdot{\bf \hat r}_{ij} -
{\bf e}_{j}\cdot{\bf \hat r}_{ij})^{2}}{1 - \chi ^{\prime}({\bf e}_{i} \cdot
{\bf e}_{j})}\right]
\end{equation}
with $\chi^{\prime} = (\kappa^{\prime ~ 1/\mu} - 1) /
(\kappa^{\prime ~ 1/\mu} + 1)$. Here
$\kappa = \sigma_{ff}/\sigma_{ee}$ is the aspect ratio of the ellipsoid of
revolution with $\sigma_{ee}$ denoting the separation between two ellipsoids of 
revolution in a edge-to-edge configuration, and $\sigma_{ee} = \sigma_{0}$, and
$\kappa^{\prime} = \epsilon_{ee}/\epsilon_{ff}$, where $\epsilon_{ee}$ is the
depth of the minimum of the potential for a pair of ellipsoids of revolution
aligned parallel in a edge-to-edge configuration, and $\epsilon_{ff}$ is the
corresponding depth for the face-to-face alignment. Here $\epsilon_{0}$ denotes  
depth of the minimum of the pair potential for cross allignment. The parameterization, that we
have employed here, is $(\kappa = 0.345, \kappa^{\prime} = 0.2, \mu = 1, \nu =2)$
\cite{Bates-Luckhurst-JCP-1996}.

Molecular dynamics simulations have been performed with the model discotic system 
containing $500$ oblate ellipsoids of revolution in a cubic box with periodic 
boundary conditions. All the quantities reported here are given in reduced units, 
defined in terms of the Gay-Berne potential parameters $\epsilon_{0}$ and $\sigma_{0}$, 
each of which has been taken to be unity: length in units of $\sigma_{0}$, temperature 
in units of $\epsilon_{0}/k_{B}$, $k_{B}$ being the Boltzmann constant, and time in 
units of $(\sigma_{0}^{2}m/\epsilon_{0})^{1/2}$, $m$ being the mass of the ellipsoids 
of revolution. We have set the mass as well as the moment of inertia of each of the 
ellipsoids of revolution equal to unity. The intermolecular potential has been 
truncated at a distance $r_{cut} = 1.6$ as in Ref. \cite{Bates-Luckhurst-JCP-1996} and 
shifted. The equations of motion have been integrated following the velocity-Verlet 
algorithm with the integration time steps of $\delta t = 0.0015$ in the reduced units 
\cite{Ilnytskyi-CPC-2002}. Equilibration has been done in an NPT ensemble for a time 
period of $t_{q}$. Following this, the system has been allowed to propagate with a 
constant energy and density for a time period of $t_{e} (\geq t_{q})$ in order to 
ensure equilibration. Upon observation of no drift in temperature, pressure, and 
potential energy, the data collection has been executed in a microcanonical 
ensemble. The model discotic system has been melted from an initial fcc 
configuration at high temperatures and low densities, and studied along two 
isobars at pressures $P = 25$ and $P = 10$ at several temperatures. 

\section{Results and discussion}

\begin{figure}
\epsfig{file=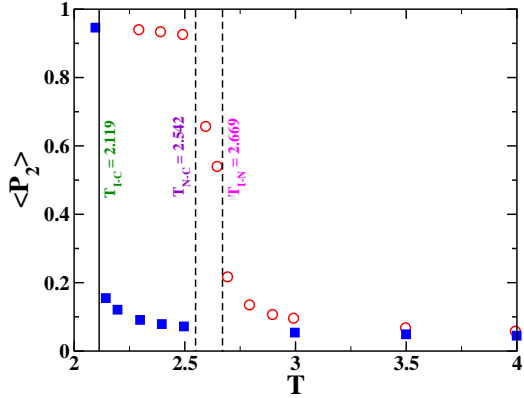,width=6.5cm,angle=270}
\caption{The average second-rank orientational order parameter $<P_{2}>$ as a 
function of temperature along two isobars. The open circles (red) correspond to the data for
the pressure $P = 25$ and the squares (blue) for $P = 10$. The phase boundaries are shown
by vertical dotted lines for $P = 25$ ($T_{I-N} = 2.669$ and $T_{N-C} = 2.542$) 
and by a vertical solid line for $P =10$ ($T_{I-C} = 2.119$).}
\label{fig:tmp-op}
\end{figure}

We first need to characterize the phases that appear along the isobars studied here.
To this end, we have monitored the average second-rank orientational order parameter
$<P_{2}>$ and the radial distribution function (data not shown here). In Fig. 
\ref{fig:tmp-op}, we show the evolution of $<P_{2}>$ with temperature along the two 
isobars. The second-rank orientational order parameter has been computed as the largest 
eigenvalue of the order parameter tensor
\begin{equation}
S_{\alpha\beta} = \frac{1}{N}\sum_{i=1}^{N}\frac{1}{2}
(3e_{i\alpha}e_{i\beta} - \delta_{\alpha\beta}),
\label{eq:defS}
\end{equation}
where $\alpha, \beta = x, y, z$ are the indices referring to the space fixed
frame, $e_{i\alpha}$ is the $\alpha$-component of the unit vector ${\bf e}_{i}$,
$\delta_{\alpha\beta}$ is the Kronecker symbol, and $N$ is the number of
ellipsoids of revolution present in the system. $<P_{2}>$ tends to zero in the
isotropic phase but retains a non-zero value because of the finite size of the
system. In the nematic phase, $<P_{2}>$ has a value above $0.4$. For the columnar
phase, $<P_{2}>$ is above $0.9$. In the present case, we observe the I-N-C phase
sequence along the isobar at the higher pressure and the sequence I-C along the
other isobar. Note the sharp jump in the $<P_{2}>$ for the I-C phase transition.

\begin{figure}
\epsfig{file=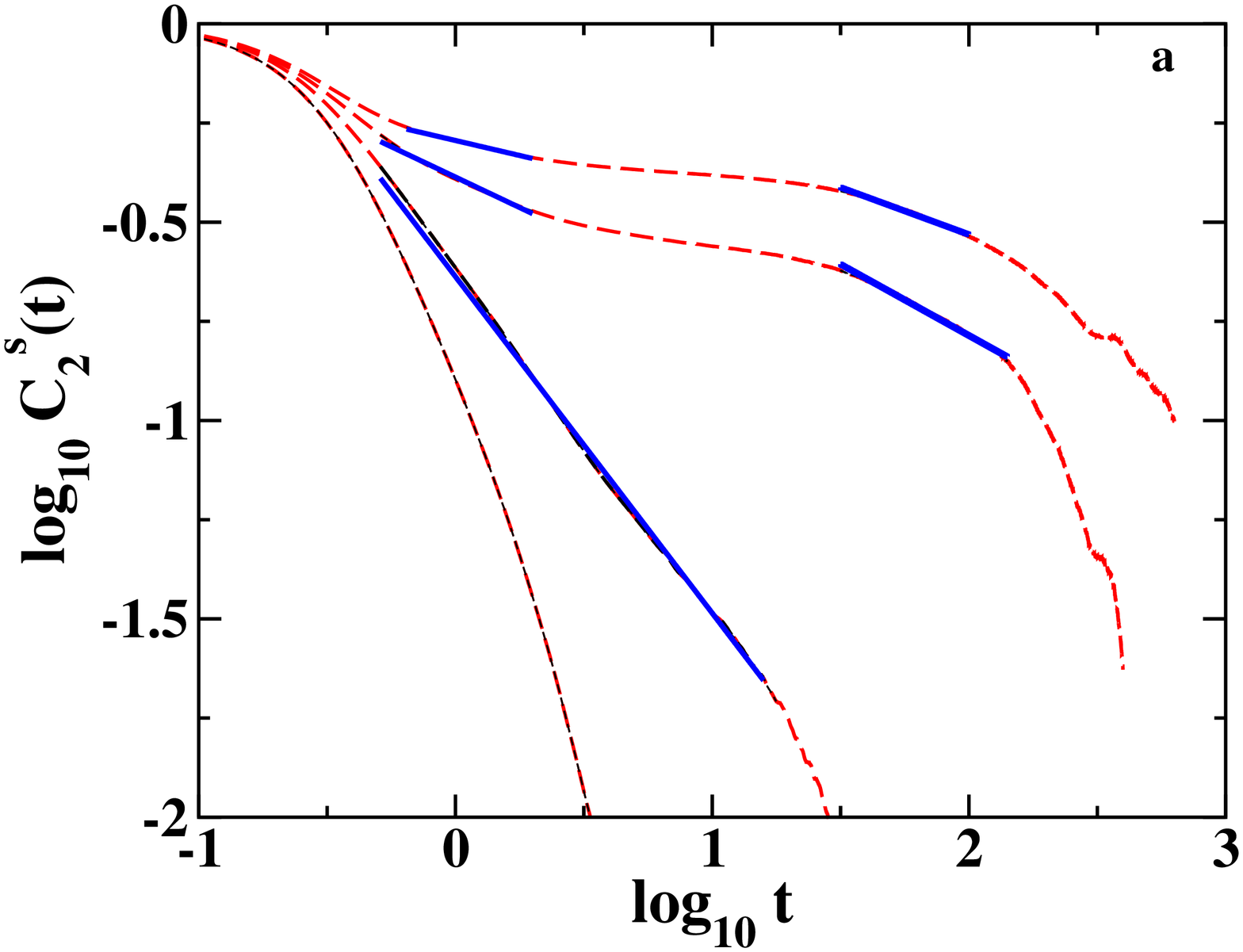,width=6.5cm,angle=270}
\epsfig{file=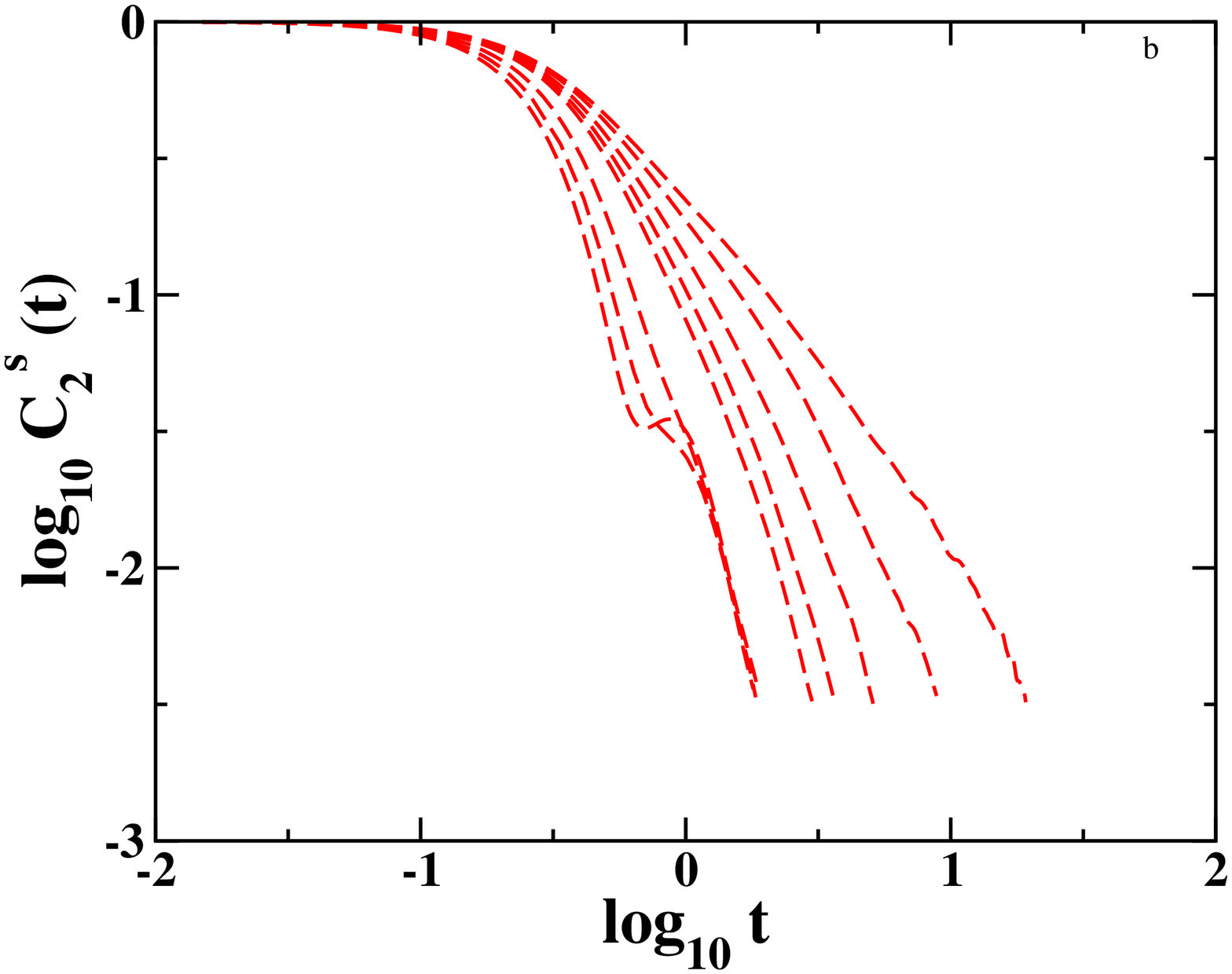,width=6.5cm,angle=270}
\caption{Time evolution of the single-particle second-rank OTCF in log-log plots 
for the discotic system at several temperatures. The dashed lines (red) are the 
simulation data corresponding to increasing orientational order parameter or decreasing 
temperature from the bottom to the top. The solid lines (blue) are the linear fits to the data, showing the 
power law decay regimes: $~ t^{-\alpha}$. The valuse of the power law exponent $\alpha$ are given 
below in the paranthesis. (a) Along the isobar at $P = 25.0$ at several 
temperatures: $ T = 2.991, 2.693 (\alpha = 0.85), 2.646 (\alpha = 0.31, 0.35)$, and $2.594 
(\alpha = 0.15, 0.23)$; (b) Along the isobar at $P = 10.0$ at all the temperatures studied 
for the isotropic phase ($ T = 3.999, 3.499, 2.997, 2.496, 2.396, 2.298, 2.196, 2.143 $). The 
power law decay regions is absent in the OTCF of isobar at $P = 10.0$.}
\label{fig:otcf2}
\end{figure}

We have investigated orientational dynamics at the single-particle level by 
monitoring the temporal evolution of the corresponding second-rank orientational 
time correlation functions (OTCF), that is defined by

\begin{equation}
C_{2}^{s}(t)={\frac{\langle{\displaystyle \sum_{i} P_{2}({\bf e}_{i}(0)\cdot
{\bf e}_{i}(t))}\rangle}{
\langle{\displaystyle \sum_i P_{2}({\bf e}_{i}(0)\cdot {\bf e}_{i}(0))}
\rangle}},
\label{eq:defotcfs}
\end{equation}

where $P_{2}$ is the second rank Legendre polynomial, and the angular 
brackets stand for ensemble averaging. In Fig. \ref{fig:otcf2}, we show the time 
evolution of the single-particle second-rank OTCF at several temperatures in 
log-log plots. The emergence of a power law decay at short-to-intermediate times 
near the I-N phase boundary is notable in Fig. \ref{fig:otcf2}(a) from the linear fit. 
It follows from Fig. \ref{fig:otcf2}(a) that as the system transits across the I-N phase 
boundary, two power law relaxation regimes, separated by a plateau, appear giving 
rise to a step-like feature. However, the decay of the single-particle second-rank
OTCF in the isotropic phase near the isotropic-columnar phase boundary {\it does 
not} follow any power law as evident in Fig. \ref{fig:otcf2}(b). Note that the step like relaxation 
feature which is observed for the former is also absent in the later. 

In optical heterodyne detected optical Kerr effect measurements (OHD-OKE), one 
probes collective orientational relaxation \cite{Torre-PRE-1998}. In recent OHD-OKE 
experiments with calamitic liquid crystals, the decay of the OKE signal has been 
found to follow a complex pattern \cite{Gottkea-JCP-2002,Gottkeb-JCP-2002,Cang-CPL-2002,Li-JPCB-2005}. 
The most intriguing feature has been the power law decay regimes at short-to-intermediate 
times \cite{Gottkeb-JCP-2002,Cang-CPL-2002}. We have therefore monitored the time 
evolution of the collective second-rank OTCF, defined by

\begin{equation}
C_{2}^{c}(t)={\frac{\langle{\displaystyle \sum_{i} \sum_{j}P_{2}({\bf e}_{i}
(0)\cdot{\bf e}_{j}(t))}\rangle}{\langle{\displaystyle \sum_{i} \sum_{j}
P_{2}({\bf e}_{i}(0)\cdot {\bf e}_{j}(0))}\rangle}}.
\label{eq:defotcfc}
\end{equation}

\begin{figure}
\epsfig{file=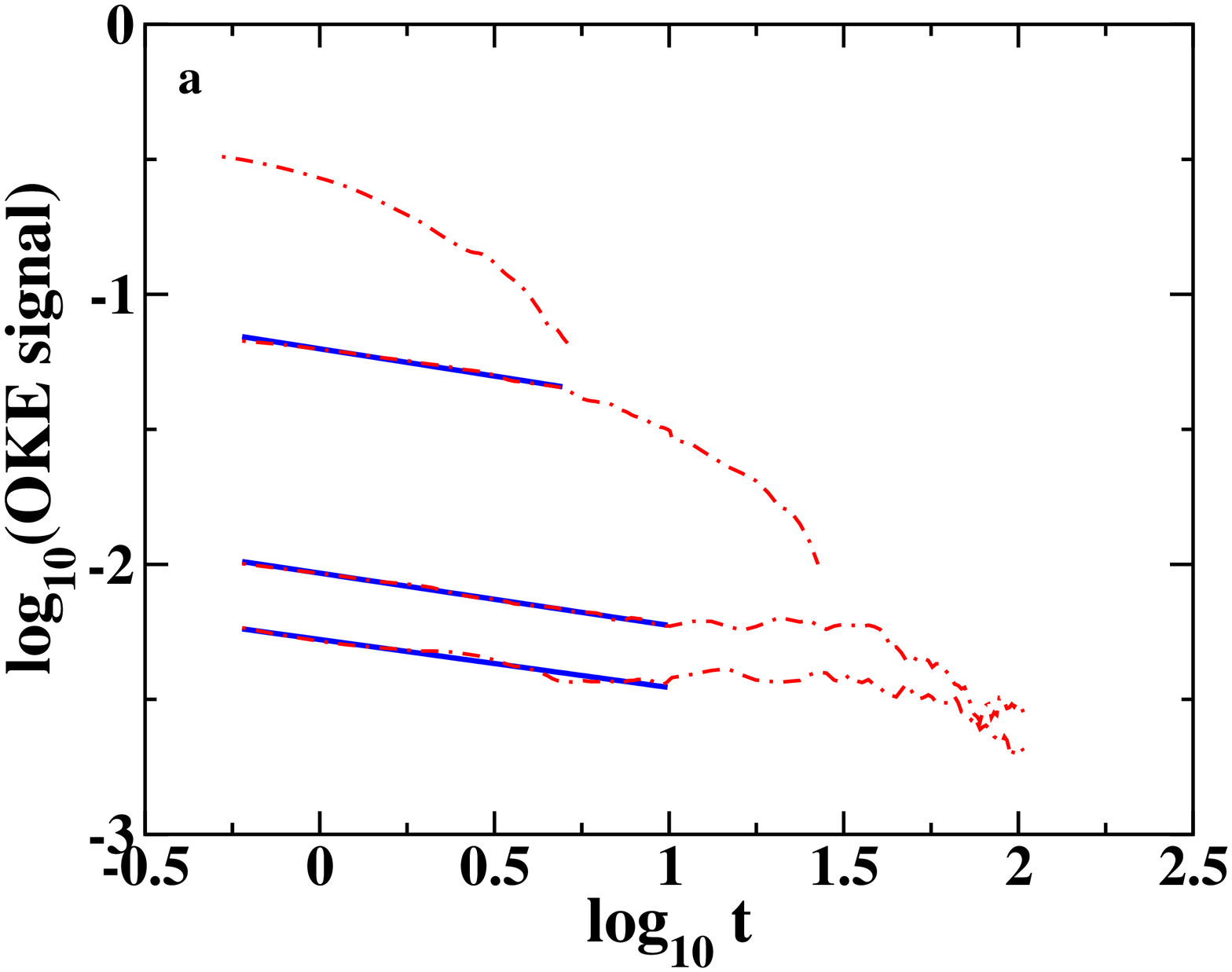,width=6.5cm,angle=270}
\epsfig{file=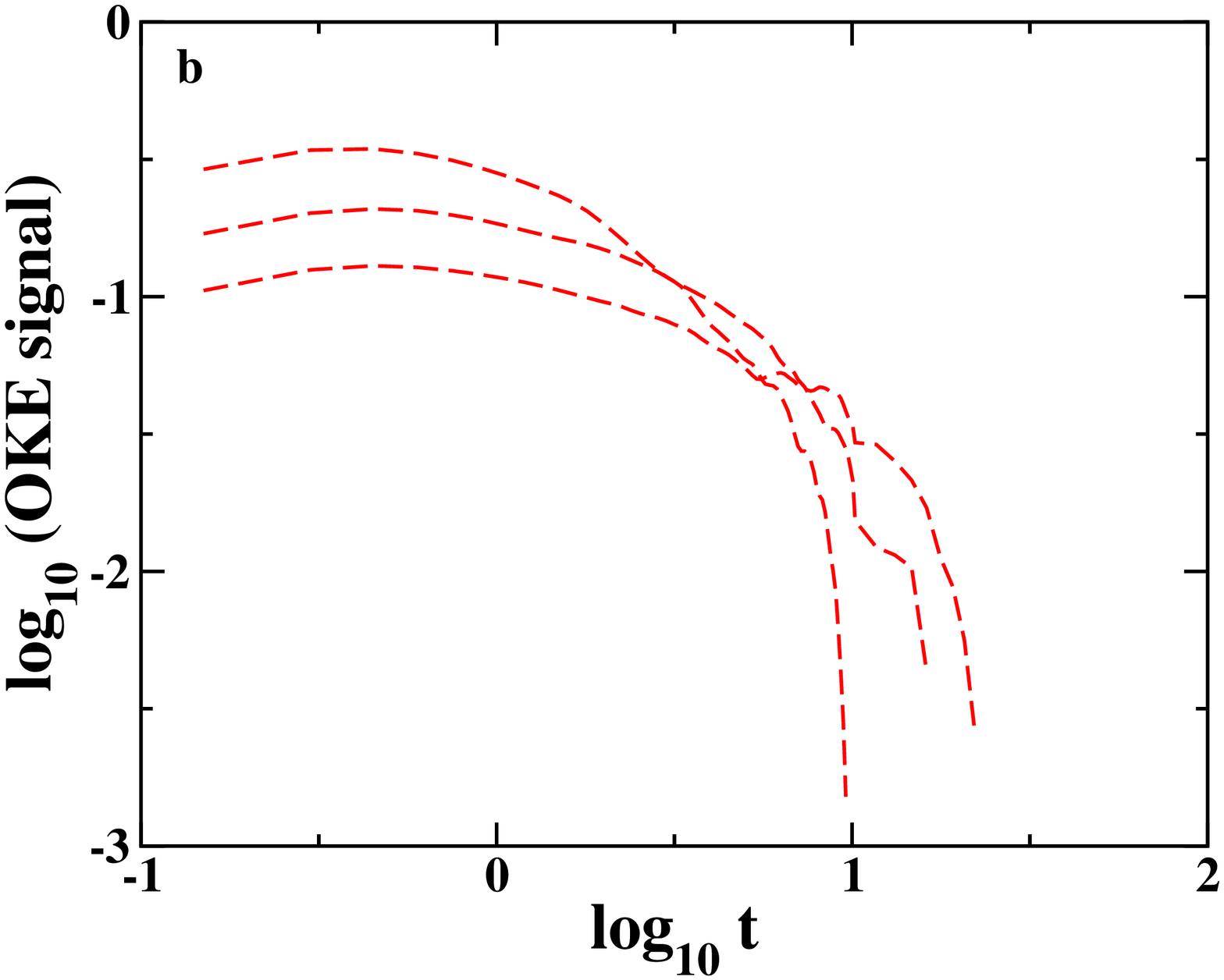,width=6.5cm,angle=270}
\caption{The short-to-intermediate time decay of the OKE signal in log-log plots 
for the discotic system. The dashed lines (red) are the simulation data and the solid 
lines show the linear fits (blue) to the data showing the power law decay regimes: 
$~ t^{-\alpha}$. The values of the power law exponent $\alpha$ are given below in 
the parenthesis. (a) Along the isobar at $P = 25.0$ at several temperatures: 
$T = 2.991$, $T = 2.693$ ($\alpha = 0.208$), $T = 2.646$ ($\alpha = 0.194$), and 
$T = 2.594$ ($\alpha = 0.178$). (b) Along the isobar at $P = 10.0$ at several 
temperatures: $T = 2.298, 2.196,$ and $2.143$ amd power law decay is absent. 
Temperature decreases from the top to the bottom at the left of the figure in each case.}
\label{fig:oke}
\end{figure}

In the present case, the negative of the time derivative of the collective
second-rank OTCF provides a measure of the experimentally observable OHD-OKE
signal. As monitoring the time evolution of $C_{2}^{c}(t)$ is computationally 
quite demanding, we have restricted ourselves to the short-to-intermediate time 
dynamics that would suffice to compare the most intriguing aspect of the 
experimental observations \cite{Gottkeb-JCP-2002,Cang-CPL-2002,Li-JPCB-2005}.
In Fig. \ref{fig:oke}, we show in log-log plots the temporal behavior of the OKE 
signal derived from present system at several temperatures. {\it A
short-to-intermediate-time power law regime} is evident in the decay of the OKE 
signal on either side of the I-N transition as illustrated by the linear fitting in
Fig. \ref{fig:oke}(a). In consistency with the single-particle dynamics, such a 
power law decay regime is {\it not} observed for the OKE signal in the isotropic 
phase near the I-C phase boundary as apparent in Fig. \ref{fig:oke}(b). 

It follows from the time evolution of the single-particle second-rank OTCF shown
in Fig. \ref{fig:otcf2}(a) that as the system settles into the nematic phase, two 
power law decay regimes, that are separated by a plateau, emerge. Such a feature 
bears a close resemblance with what has been observed recently for a model system
of calamitic liquid crystals \cite{Chakrabarti-PRL-2005}. The decay pattern is also
similar to those observed for models supercooled molecular liquids 
\cite{Kammerer-PRE-1997,Michele-PRE-2001}. In fact, based on a series of OHD-OKE 
measurements Fayers and coworkers have recently drawn an analogy in the orientational 
dynamics between calamitic liquid crystals in their isotropic phase near the I-N 
transition and supercooled molecular liquids \cite{Cang-JCP-2003}. The analogous 
dynamics could be captured in a subsequent molecular dynamics simulation study of 
model systems of these two classes of soft condensed matter 
\cite{Chakrabarti-Bagchi-Unpublished}. The short-to-intermediate time power law decay 
of the OKE signal observed therein bears a close similarity with what is found in
the present discotic system across the I-N transition.   
 
The contrasting behavior observed in orientational relaxation in the isotropic phase 
near the I-N and the I-C phase boundaries is noteworthy. Such an observation may 
throw new light on the origin of the power law relaxation in the isotropic phase near 
the I-N transition. While the I-C transition is strongly first order in nature, the I-N 
transition is only weakly first order with certain characteristics of the continuous 
transition. This is reflected in the present case in a much larger change in the density
marking the I-C transition as compared to the I-N transition (data not shown). The weakly 
first order nature of the I-N transition appears to play a role in the short-to-intermediate 
time power law relaxation. It seems fair to trace the origin of the power law decay in 
orientational relaxation to the growth in the orientational correlation length in the 
isotropic phase near the I-N transition. To this end, we attempt a theoretical analysis in 
the next section.

\section{Theoretical analysis}

The I-N phase transition is weakly first order both in calamitic and discotic systems. This 
is manifested in the growing orientational pair correlation length as the I-N phase boundary 
is approached from the high temperature isotropic phase. Apparently, a second order phase 
transition at a temperature only slightly lower (by $\sim 1 K$), where the orientational
correlation length would have diverged, is preempted by the weakly first order phase 
transition. Nevertheless, even this weakly first order phase transition is driven by the 
growing correlation length. The temperature dependent growth of this correlation length 
$\xi (T)$ can be given by the following expression \cite{deGennes-book}
i
\begin{equation}
\xi (T) = A (T^{*} - T)^{-\nu}
\end{equation}

where $\nu$ is $0.5$ in the Landau mean-field theory.

A simple mode coupling theory, based on time dependent density functional theory, 
shows that this growing correlation length can give rise to a power-law decay of
the type observed in simulations. This approach uses the the generalized
Debye-Stokes-Einstein relation between the correlation time, diffusion, and
friction \cite{sarika-ACP-2001}
\begin{equation}                                                                   
C_{2}(z) = \frac{1}{(z+ 6AD_{R} (z))} 
\end{equation}
and
\begin{equation}
D_{R}(z) = \frac{k_{B}T}{I(z + \zeta(z))}, 
\end{equation}
where $A$ is equal to 1 for the single-particle relaxation, but is related to 
orientational caging for collective dynamics. It was shown elsewhere, the growing 
correlation length can give rise to a singular frequency dependence of $\zeta$ over a 
frequency range $\zeta(z) \thicksim A/z^{\alpha}$ with $\alpha = 0.5$ \cite{Gottkea-JCP-2002,
Gottkeb-JCP-2002}. This power law dependence in the frequency dependence of friction in turn gives 
rise to a power law decay in the orientational time correlation function along with the slowing down of 
the relaxation.

Thus, in the above mentioned theory, the origin of the power law decay is 
essentially the same as observed near the critical phenomena \cite{Kunter-PRB-1982}. However, 
one may not expect a universal behavior since there is no true divergence. The 
absence of power law decay near the I-C phase boundary could then be due to the 
absence of any growing correlation length. The I-C phase transition is strongly 
first order in nature where both orientational and positional order set in at 
the same time. Since the growth of orientational correlation is small, a power 
law decay is not expected.  

\section{Orientational Pair Distribution}

To this end, we have calculated the distance 
dependent orientational pair distribution function $g_{ll^{\prime}m}(r)$ 
\cite{Bates-JCP-1999,Allen-Tildesley-Book} wihich is defined by

\begin{equation}
g_{ll^{\prime}m}(r) = 4\pi g(r)\langle{Y_{lm}^{\star}(\omega_{i})Y_{l^{\prime}m}^{\star}
(\omega_{j})\rangle}      
\end{equation}
Here, $Y_{lm}^{\star}(\omega)$ is a spherical hermonics and $\omega$ denotes the sperical 
polar angle made by the particle symmetry axis with the intermolecular separation vector. 
it can be shown that at sufficiently large separation at which orientations are uncorrelated, 
two spherical harmonics can be averaged independently of each other. When the director is taken to define 
$z$ axis of the laboratory frame, separate averaging gives

\begin{equation}
\lim_{r\rightarrow \infty}g_{ll^{\prime}m}(r) = (-1)^{m} g(r) \delta_{l,l^{\prime}} \langle{
P_{l}\rangle} \langle{P_{l^{\prime}}\rangle} 
\end{equation}

\begin{figure}
\epsfig{file=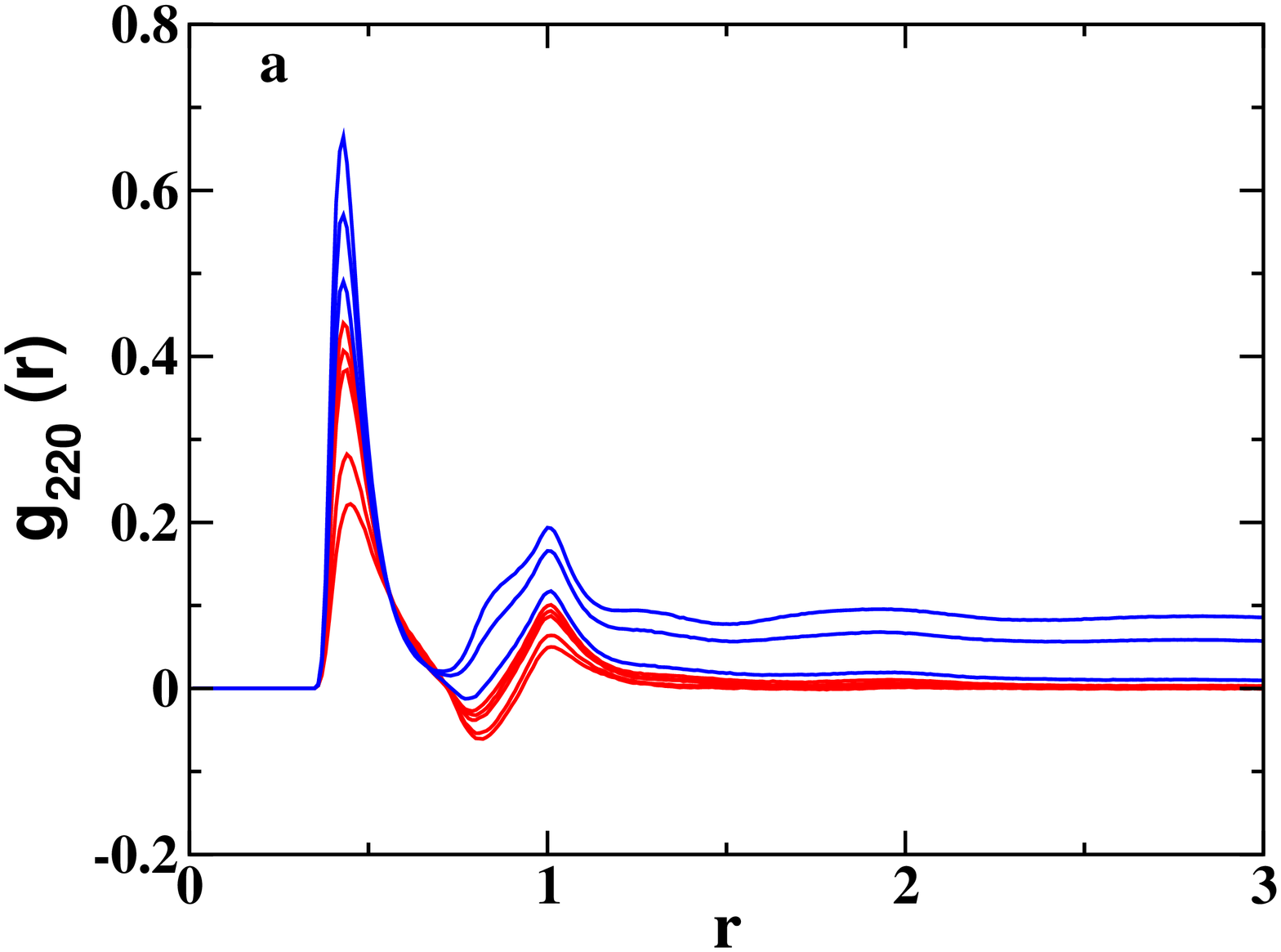,width=6.5cm,angle=270}
\epsfig{file=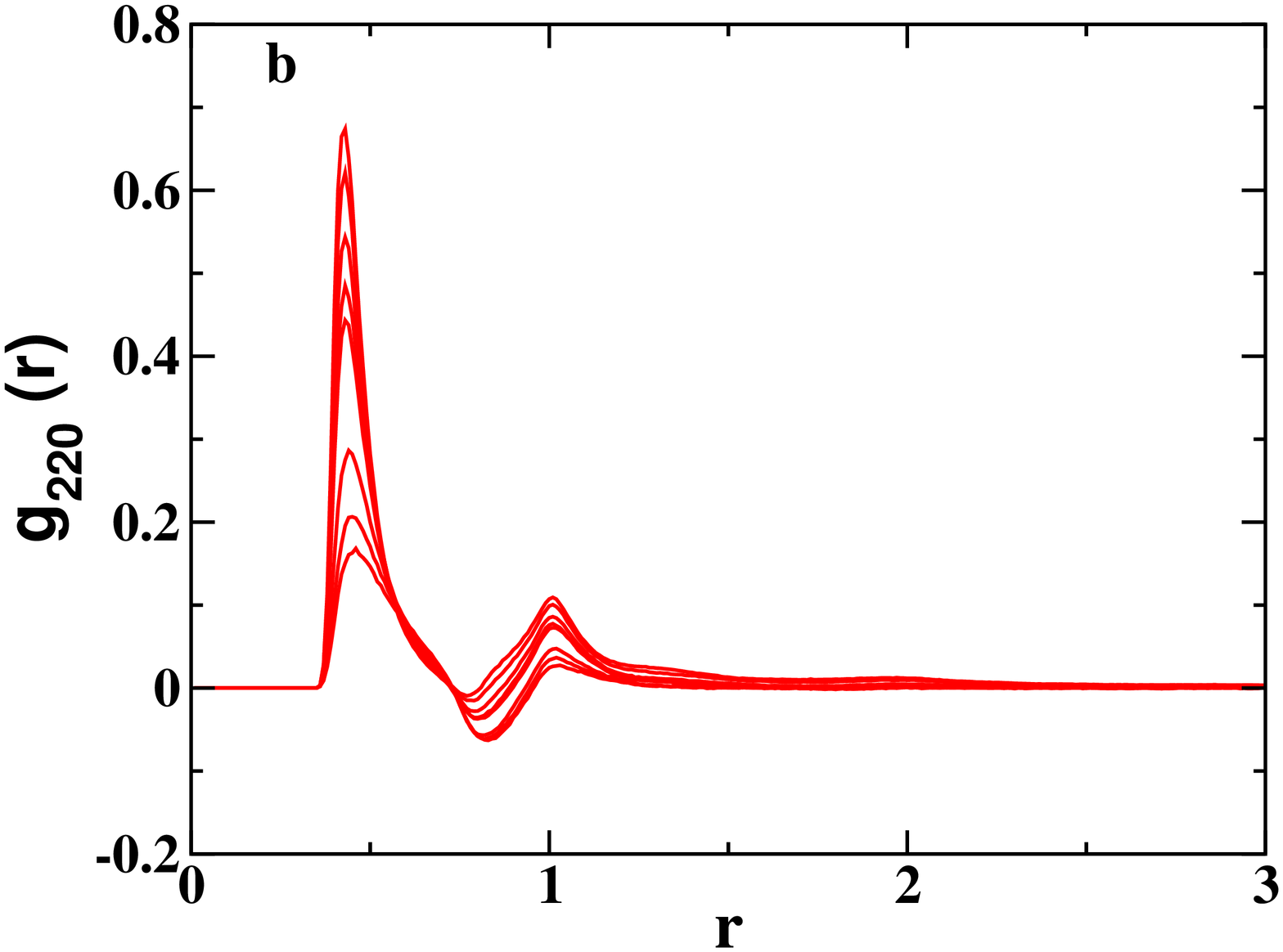,width=6.5cm,angle=270}
\caption{The orientational pair distribution function $g_{220} (r)$ for the model
discotic system along the two isobars: (a) the one at $P = 25$ and (b) the
other at $P = 10$. The temperature decreases from the bottom to the top at the position
of the dominant peak of the curves starting from high temperature isotropic phase down
to the temperature which is just above the temperature at which columnar phase appears. The 
growth of long range orientational correlation is notable from the blue curves for I-N-C isobar.}
\label{fig:gr}
\end{figure}

This coefficient will be vanished at large separations when ranks $l$ and $l^{\prime}$ are different. 
While for $l=l^{\prime}$ in the isotropic phase where order parameter vanishes and the coefficient 
$g_{ll^{\prime}m}(r)$ tend to zero for large separatios, it will have a finite value if long range 
orientational correlation is devloped in the system. To verify our assertion made in theoretical section,
we have calculated $g_{220}(r)$ for the systems studied here along both the isobars and presented in Fig. 
\ref {fig:gr}(a) and Fig. \ref {fig:gr}(b), respectively. While the growth of orientational correlation 
length is clearly evident across the I-N transition for large separations, such a growth is found to be 
totally absent in the isotropic phase near the I-C phase boundary.

\section{Conclusion}

Let us first summarize the main results of the present work. In order to understand
orientational relaxation in disk-like molecules that form discotic phase on cooling, 
we have performed molecular dynamics
simulations of a model system that consists of oblate
ellipsoids of revolution interacting with each other via a variant of the Gay-Berne pair 
potential. The system has been studied along two isobars so chosen that the phase 
sequence I-N-C appears upon cooling along the one and the sequence I-C along the 
other. We have investigated temperature dependent orientational relaxation across 
the I-N transition and in the isotropic phase near the I-C phase boundary with a 
focus on the short-to-intermediate time decay behavior. While the orientational 
relaxation across the I-N phase boundary shows a power law decay at 
short-to-intermediate times, such power law relaxation is not observed in the 
isotropic phase near the I-C phase boundary. Study of orientational pair distribution 
function shows that there is a growth of orientational pair correlation near the 
I-N transition whereas such a growth is absent in the isotropic phase near the I-C 
phase boundary. As the system settles into the nematic phase, the decay of the single-particle 
second-rank orientational time correlation function follows a pattern that is similar to 
what is observed with calamitic liquid crystals and supercooled molecular liquids 
\cite{Chakrabarti-PRL-2005,Kammerer-PRE-1997,Michele-PRE-2001}.

In order to further understand microscopic slowing down of the collective OTCF, 
we recall the expression \cite{Kivelson-ARPC-1980} 
\begin{equation}
\label{collective-single}
\frac{\tau_{2}^{c}}{\tau_{2}^{s}} = \frac{(1 + g_{2})}{(1 + j_{2})}
\end{equation}
where $g_{2}$ is the static second rank Kirkwood factor \cite{Allen-PRL-1987}
\begin{equation}
\label{kirkwood}
g_{2} = \sum_{j \neq i} P_{2}(e_{i}\cdot e_{j}) = \frac{1}{N} \sum_{i}\sum_{j \neq
i}P_{2}(e_{i}\cdot e_{j})
\end{equation}
and $j_{2}$ is a dynamic quantity which can be expressed in terms memory functions of orientation as 
\cite{Bern-Pecora-DynamicLightScattering-1975}
\begin{equation}
j_{2} = N \frac {\int_{0}^{\infty} \langle{\dot \alpha^{yz}_{1} e^{iQLt}\dot \alpha^{yz}_{2}
\rangle}}{\int_{0}^{\infty} \langle{\dot \alpha^{yz}_{1} e^{iQLt}\dot \alpha^{yz}_{1}\rangle}}
\end{equation}
where $\dot \alpha^{yz}_{1} = (\alpha_{\parallel} - \alpha_{\perp}) iL(e^{y}_{1}e^{z}_{1})$ and $Q$ 
is the projection operator. Here $L$ is the Liouville operator and $e^{x,y}_{i}$ is the $x (or~y)$ component 
of the unit vector along the short axis of the $i^{th}$ oblate ellipsoid of revolution. It is non-trivial 
to calculate $j_{2}$ from first principles but one can always estimate it from Eq \ref{collective-single} 
where other qnatities are not hard to calculate in principle.

 We have calculated static second rank Kirkwood $g_{2}$ 
factor \cite{anonymous-referee} which may be thaught of the average number of molecules 
whose orientations are perfectly correlated to that of a given molecule \cite{Cheung-JCSFT-1997}. 
We find that $g_{2}$ shows the same behaviour as the order parameter variation for both the isobars studied 
here. Dynamic quantity $j_{2}$ has no such straight foroward physical interpretation like 
$g_{2}$. Because of the slow power law decay, it has not been possible to calculate the relaxation times 
($\tau_{2}^{c}$ and $\tau_{2}^{s}$) near the I-N phase boundary. However, when calculated away 
from the phase boundary, when nrelaxation functions are nearly single exponential, the value of
$j_{2}$ is found to be small.  The dynamic quantity usually has small and negative value and also 
will not have much variation across the transitions as observed for several studies 
\cite{Allen-PRL-1987,Alms-JCP-1973}.

In contrast to our observation of the lack of power law decay in orientational
relaxation of the discotic system in the isotropic phase near the I-C phase boundary, a 
very recent OHD-OKE experimental study by Fayer and coworkers finds a power law $t^{-0.76}$ at 
short times and von schweidler power law $t^{-0.26}$ at intermediate times along with a long time 
exponential relaxation in the isotropic phase above the I-C transition \cite{Li-JCP-2006}. 
The decay pattern is somewhat similar to what was observed for the calamitic system in the 
isotropic phase near I-N phase transition. It is possible that nematic fluctuations were 
important in their experimental system depending upon the choice of temperature. This point 
deserves further study.

\begin{center}
{\large \bf Acknowledgments}
\end{center}

This work was supported in part by the grants from the DST, India and the CSIR, India. DC 
acknowledges UGC, India and BJ acknowledges CSIR, India for providing financial support.


\begin{references}

\bibitem{deGennes-book} P. G. de Gennes and J. Prost, {\it The Physics of 
Liquid Crystals}, (Clarendon Press, Oxford, 1993).

\bibitem{Chandrasekhar-book} S. Chandrasekhar, {\it Liquid Crystals} (Cambridge
University Press, Cambridge, 1992).

\bibitem{Chandrasekhar-Pramana-1970} S. Chandrasekhar, B. K. Sadashiva, and K. A. 
Suresh, Pramana {\bf 9}, 471 (1977).

\bibitem{Ohta-Proc-1999} K. Ohta {\it et al.}, Proc. $13^{th}$. Inter. Conf.
Dielec. Liq. (IDCL '99), Nara, Japan, {\bf July 20-25}, 561 (1999).

\bibitem{Adam-Nature-1994} D. Adam {\it et al.}, Nature. {\bf 371}, 141 (1994).

\bibitem{Yoshino-Proc-1999} K. Yoshino {\it et al.}, Proc. $13^{th}$. Inter. Conf. 
Dielec. Liq. (IDCL '99), Nara, Japan, {\bf July 20-25}, 598 (1999).  

\bibitem{Chandrasekhar-Ranganath-RPP-1990} S. Chandrasekhar and G. S. Ranganath, 
Rep. Prog. Phys. {\bf 53}, 57 (1990).

\bibitem{Hindmarsh-JMC-1993} P. Hindmarsh, M. Hird, P. Styring, and J. W. Goodby,
J. Mater. Chem. {\bf 3}, 1117 (1993).

\bibitem{Pasini-Zannoni-book} P. Pasini and C. Zannoni, eds., {\it Advances in
the Computer Simulations of Liquid Crystals}, (Kluwer Academic Publishers, 
Dordrecht, 2000).

\bibitem{Wilson-IRPC-2005} M. R. Wilson, Int. Rev. Phys. Chem. {\bf 24}, 421 
(2005).

\bibitem{Zannoni-JMC-2001} C. Zannoni, J. Mater. Chem. {\bf 11}, 2637 (2001).

\bibitem{Allen-ACP-1993} M. P. Allen, G. T. Evans, D. Frenkel, and B. M. Mulder, 
Adv. Chem. Phys. {\bf 86}, 1 (1993).

\bibitem{Andersen-ACP-1976} H. C. Andersen, D. Chandler, and J. D. Weeks, Adv.
Chem. Phys. {\bf 34}, 105 (1976).

\bibitem{Eppenga-Frenkel-MP-1984} R. Eppenga and D. Frenkel, Mol. Phys. {\bf 52}, 
1303 (1984).

\bibitem{Frenkel-Mulder-MP-1985} D. Frenkel and B. M. Mulder, Mol. Phys. {\bf 55},
1171 (1985).

\bibitem{Allen-PRL-1990} M. P. Allen, Phys. Rev. Lett. {\bf 65}, 2881 (1990).

\bibitem{Veerman-Frenkel-PRA-1992} J. A. C. Veerman and D. Frenkel, Phys. Rev. A
{\bf 45}, 5632 (1992).
  
\bibitem{Gay-Berne-JCP-1981} J. G. Gay and B. J. Berne, J. Chem. Phys. {\bf 74}, 
3316 (1981).

\bibitem{Emerson-MP-1994} A. P. J. Emerson, G. R. Luckhurst, and S. G. Whatling, 
Mol. Phys. {\bf 82}, 113 (1994).

\bibitem{Bates-Luckhurst-JCP-1996} M. A. Bates and G. R. Luckhurst, J. Chem. Phys.
{\bf 104}, 6696 (1996).

\bibitem{Capiron-PRE-2003} D. Caprion, L. Bellier-Castella, and J.-P. Ryckaert,
Phys. Rev. E {\bf 67}, 041703 (2003).

\bibitem{Coussaert-Baus-JCP-2002} T. Coussaert and M. Baus, J. Chem. Phys. 
{\bf 116}, 7744 (2002).

\bibitem{Maliniak-JCP-1992} A. Maliniak, J. Chem. Phys. {\bf 96}, 2306 (1992).

\bibitem{Zamir-JACS-1994} S. Zamir {\it et al.}, J. Am. Chem. Soc. {\bf 116}, 1973
(1994).

\bibitem{Groothues-LC-1995} H. Groothues, F. Kremer, D. M. Collard, C. P. Lillya,
Liq. Cryst. {\bf 18}, 117 (1995).

\bibitem{Dong-MP-1999} R. Y. Dong, N. Boden, R. J. Bushby, and P. S. Martin, Mol.
Phys. {\bf 97}, 1165 (1999).

\bibitem{Dong-Morcombe-LC-2000} R. Y. Dong and C. R. Morcombe, Liq. Cryst. 
{\bf 27}, 897 (2000).

\bibitem{Dvinskikh-PRE-2002} S. V. Dvinskikh, I. Fur\'{o}, H. Zimmermann, and A.
Maliniak, Phys. Rev. E {\bf 65}, 050702(R) (2002).

\bibitem{Mulder-JACS-2003} F. M. Mulder {\it et al.}, J. Am. Chem. Soc. {\bf 125},
3860 (2003).

\bibitem{Zhang-Dong-PRE-2006} J. Zhang and R. Y. Dong, Phys. Rev. E {\bf 73},
061704 (2006).

\bibitem{Chakrabarti-PRL-2005} D. Chakrabarti, P. P. Jose, S. Chakrabarty, and B.
Bagchi, Phys. Rev. Lett. {\bf 95}, 197801 (2005).
 
\bibitem{Gottkea-JCP-2002} S. D. Gottke, D. D. Brace, H. Cang, B. Bagchi, and M. D. Fayer, J. 
Chem. Phys. {\bf 116}, 360 (2002).

\bibitem{Gottkeb-JCP-2002} S. D. Gottke, H. Cang, B. Bagchi, and M. D. Fayer, J. 
Chem. Phys. {\bf 116}, 6339 (2002).

\bibitem{Cang-CPL-2002} H. Cang, J. Li, M. D. Fayer, Chem. Phys. Lett. {\bf 366},
82 (2002).

\bibitem{Li-JPCB-2005} J. Li, I. Wang, and M. D. Fayer, {\bf 109}, 6514 (2005).

\bibitem{Jose-Bagchi-JCP-2004} P. P. Jose and B. Bagchi, J. Chem. Phys. 
{\bf 120}, 11256 (2004).

\bibitem{Bertolini-JPCB-2005} D. Bertolini, G. Cinacchi, L. D. Gaetani, and A. 
Tani, J. Phys. Chem. B {\bf 109}, 24480 (2005).

\bibitem{Ilnytskyi-CPC-2002} J. M. Ilnytskyi, and M. R. Wilson, Comput. Phys. Comm.
{\bf 148}, 43 (2002). 

\bibitem{Torre-PRE-1998} R. Torre, P. Bartolini, and R. M. Pick, Phys. Rev. E
{\bf 57}, 1912 (1998).

\bibitem{Kammerer-PRE-1997} S. K\"{a}mmerer, W. Kob, and R. Schilling, Phys.
Rev. E {\bf 56}, 5450 (1997).

\bibitem{Michele-PRE-2001} C. D. Michele and D. Leporini, Phys. Rev. E
{\bf 63}, 36702 (2001).

\bibitem{Cang-JCP-2003} H. Cang, J. Li, V. N. Novikov, and M. D. Fayer, J. Chem.
Phys. {\bf 118}, 9303 (2003).

\bibitem{Chakrabarti-Bagchi-Unpublished} D. Chakrabarti, and B. Bagchi, (manuscript 
in preparation)

\bibitem{sarika-ACP-2001} B. Bagchi, and S. Bhattacharyya, Adv. Chem.
Phys. {\bf 116}, 67 (2003).

\bibitem{Kunter-PRB-1982} R. Kutner, K. Binder, and K. W. Kehr, Phys. Rev. B
{\bf 26}, 2967 (1982).

\bibitem{Bates-JCP-1999} M. A. Bates, and G. R. Luckhurst, J. Che. Phys. {\bf 110}, 
7087 (1999). 

\bibitem{Allen-Tildesley-Book} M. P. Allen, and D. J. Tildesley, {\it Computer simulation 
of liquids}, (Clarendon Press, Oxford, 1987). 

\bibitem{Kivelson-ARPC-1980} D. Kivelson, and P. A. Madden, Ann. Rev. Phys. Chem. 
{\bf 31}, 523 (1980).

\bibitem{Allen-PRL-1987} M. P. Allen, and D. Frenkel, Phys. Rev. Lett., {\bf 58},
1748 (1987).

\bibitem{Bern-Pecora-DynamicLightScattering-1975} B. J. Berne, and R. Pecora, {\it Dynamic 
light scattering}, (John Wiley \& sons, Inc, New York, 1975)

\bibitem{anonymous-referee} We thank anonymous referee for pointing out this issue.

\bibitem{Cheung-JCSFT-1997} T. W. Cheung, S. Fan, G. R. Luckhurst, and D. L. Turner, 
J. Chem. Soc., Faraday Trans., {\bf 93}, 3099 (1997).

\bibitem{Alms-JCP-1973} G. R. Alms, D. R. Bauer, J. I. Brauman, and R. Pecora, J. Chem. 
Phys. {\bf 59}, 5310 (1973).

\bibitem{Li-JCP-2006} J. Li, K. Fruchey, and M. D. Fayer, J. Chem. Phys. {\bf 125}, 
194901 (2006).

\end{references}
\end{document}